\documentclass{iau} 
\usepackage{graphicx}

\title[Stellar population synthesis of the Milky Way] %% give here short title %%
{New population synthesis approach: \\
the golden path to constrain 
stellar and Galactic physics}

\author[Lagarde, N. and Reyl\'e C.]   %% give here short author list %%
{Nad\`ege Lagarde$^1$
 \and C\'eline Reyl\'e$^1$}

\affiliation{$^1$Institut UTINAM, CNRS UMR 6213, Univ. Bourgogne Franche-Comt\'e, OSU THETA Franche-Comt\'e-Bourgogne, Observatoire de Besan\c con, BP 1615, 25010, Besan\c con Cedex, France \\ email: {\tt nadege.lagarde@utinam.cnrs.fr} \\}

\pubyear{2019}
\volume{IAUS 357}  %% insert here IAU Symposium No.
\jname{White Dwarfs as probes of fundamental physics and tracers of planetary, stellar \& galactic evolution}
\editors{A.C. Editor, B.D. Editor \& C.E. Editor, eds.}

\begin{document}

\maketitle

\begin{abstract}
The cornerstone mission of the European Space Agency, Gaia, has revealed properties of 260 000 white dwarfs in the Galaxy, allowing us for the first time to constrain the evolution of white dwarfs with a large sample. Complementary surveys (CoRoT, \textit{Kepler}, K2, APOGEE and Gaia-ESO), will revolutionize our understanding of the formation and history of our Galaxy, providing accurate stellar masses, radii, ages, distances, and chemical properties for very large samples of stars across different Galactic stellar populations.
To exploit the potential of the combination of spectroscopic, seismic and astrometric observations, the population synthesis approach is a very crucial and efficient tool. We develop the Besan\c con Galaxy model (BGM, Lagarde et al 2017) for which stellar evolution predictions are included, providing the global asteroseismic properties and the surface chemical abundances along the evolution of low- and intermediate-mass stars. For the first time, the BGM can explore the effects of an extra-mixing occurring in red-giant stars (Lagarde et al. 2019), changing their stellar properties. The next step is to model a consistent treatment of giant stars and their remnants (e.g., white dwarfs). This kind of improvement would help us to constrain stellar and Galactic evolutions.

\end{abstract}

\firstsection % if your document starts with a section,
\section{Observational context}

\textbf{The Gaia space mission} provided astrometry and photometry for one billion stars, with typical uncertainties of about 0.3 mas for the positions and parallaxes, and about 1 mas/yr for the proper motions (Data release 2, \cite[Gaia Collaboration 2018a]{GaiaDR2}). The last data release of Gaia has already provided precise astrometric and photometric data for 260,000 high-confidence white dwarf candidates (\cite[Gentile Fusillo \ et al.\  2019]{Gentile19}), and the future data releases should highlight $\sim$ 400,000 white dwarfs in different Galactic regions (\cite[Jordan 2007]{Jordan07}). As shown by the \cite[Gaia collaboration (2018b)]{Babusiaux18}, the incredible accuracy of Gaia data allows us to distinguish the hydrogen white dwarfs from the helium white dwarfs in the colour-magnitude diagram. %These feature has been already highlight by the SDSS data in a color-color diagrams. \\% promising a huge impact on our understanding of this kind of stars. \\

Indeed, a broad effort is ongoing with \textbf{large spectroscopic surveys} such as the Sloan Digital Sky Survey (SDSS, \cite[Kepler et al. 2016]{Kepler16}), discovering more than 30 000 white dwarfs in the data release 12 (Kepler et al. 2016). The SDSS has allowed numerous detailed spectroscopic studies of white dwarfs, specifically for DA and DB white dwarfs (e.g., \cite[Eisenstein \etal\  2006]{Eisenstein06}, \cite[Kepler \etal\  2007]{Kepler07}, \cite[Tremblay \etal\  2011]{Tremblay11}, \cite[Koester \& Kepler 2015]{KoKe15}, \cite[Genest-Beaulieu \& Bergeron 2019]{GeBe19}). \\

Finally, \textbf{asteroseismology data} of stars observed by the space missions CoRoT (\cite[Baglin \& Fridlund 2006]{CoRoT}), \textit{Kepler} (\cite[Gilliland \etal\ 2010]{Kepler}), and K2 and TESS (\cite[Ricker \etal\  2015]{TESS}) provide crucial constraints on stellar properties such as masses, radii, evolutionary states, and internal rotation profile especially for red-giant stars (e.g. \cite[Stello \etal\ 2008]{Stello08}, \cite[Mosser \etal\ 2012b]{Mosser12b}, \cite[Bedding \etal\ 2011]{Bedding11}, \cite[Vrard \etal\  2016]{Vrard16}, \cite[Mosser \etal\  2012a]{Mosser12a}, \cite[Beck \etal\ 2012]{Beck12} ). Most white dwarfs go through a pulsating stage allowing the study of their internal properties using asteroseismology (e.g., \cite[Fontaine \& Brassard 2008]{FoBr08}, \cite[Winget \& Kepler 2008]{WiKe08}, \cite[Althaus \etal\  2010]{Althaus10}, \cite[Vauclair 2013]{Vauclair13}, \cite[Corsico \etal\  2019]{Corsico19}, and more detail in Corsico's review in this volume). All of these data have been and will be collected in different regions of our Galaxy, providing a unique opportunity to constrain stellar and Galactic models. 

\section{Galactic stellar populations synthesis}

\begin{figure}[t]
%% \vspace*{-2.0 cm}
\begin{center}
 \includegraphics[width=13cm]{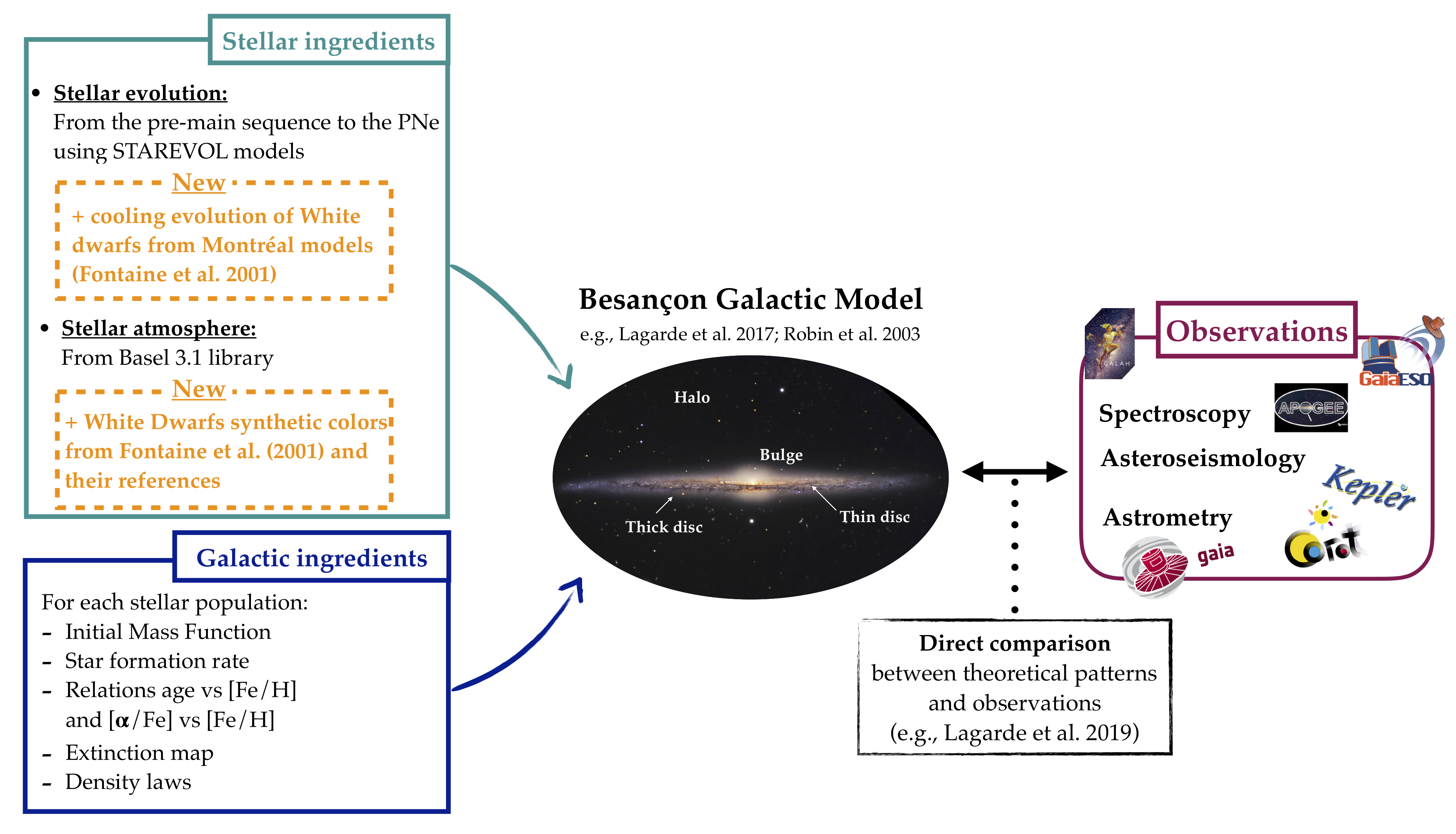} 
% %\vspace*{-1.0 cm}
 \caption{The schematic view of the Besan\c con Galactic stellar population synthesis model. The new improvements concerning the treatments of white dwarfs are shown in orange. }
   \label{fig1}
\end{center}
\end{figure}

To exploit their full potential, it is crucial to perform a combined analysis of these different kinds of observations. The population synthesis approach is a powerful tool for such analysis,
allowing the computation of mock catalogues under various model hypothesis, and to statistically compare these simulated catalogues with any type of large survey data. The stellar population synthesis approach is a useful tool to compare directly theoretical stellar and Galactic models with observations (see Fig.\ref{fig1}). The Besan\c con Galaxy model (BGM) is a stellar population synthesis model (e.g., \cite[Robin \etal\ 2003]{Robin03}, \cite[Czekaj \etal\  2014]{Czekaj14}, \cite[Lagarde \etal\ 2017]{Lagarde17}) intended to meld the formation and evolution scenarios of the Galaxy, stellar formation and evolution theory, and models of stellar atmospheres, as well as dynamical constraints, in order to make a consistent picture of the Galaxy in comparison with available observations (photometry, asteroseismology, astrometry, and spectroscopy) at different wavelengths. \\

\begin{figure}[t]
%% \vspace*{-2.0 cm}
\begin{center}
 \includegraphics[width=8cm]{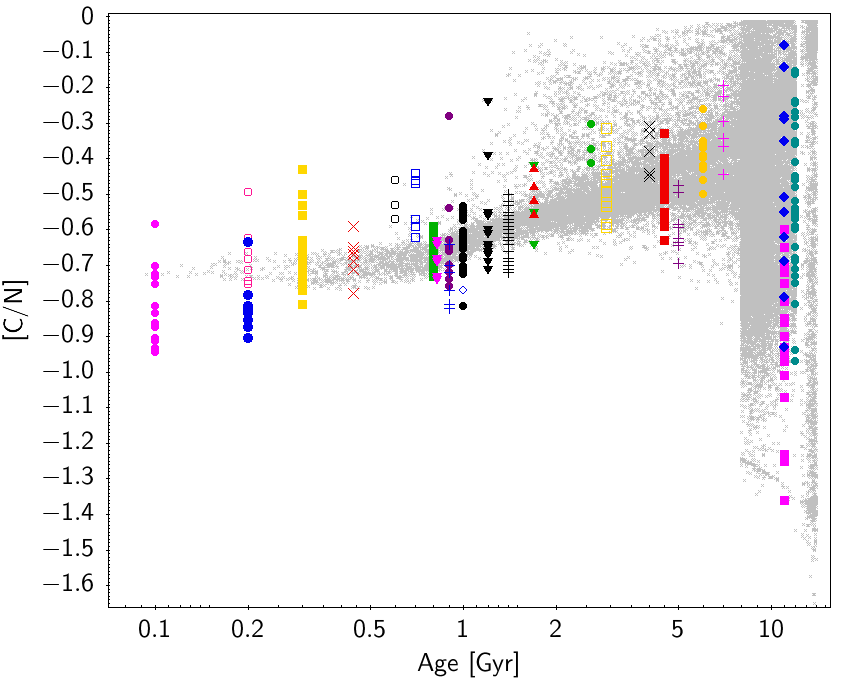} 
% %\vspace*{-1.0 cm}
 \caption{\textit{Figure from \cite[Lagarde \etal\ (2019)]{Lagarde19}} : [C/N] as a function of stellar ages for synthetic populations computed with the BGM including the effects of thermohaline instability (gray dots).The observed [C/N] abundances observed by UVES at the surface of giant stars members of open and globular clusters are also shown ((each symbol represents a cluster, see table 1 of \cite[Lagarde \etal\ (2019)]{Lagarde19}).}
   \label{fig2}
\end{center}
\end{figure}

To benefit from the combination of recent asteroseismic and spectroscopic surveys, \cite[Lagarde \etal\ (2017)]{Lagarde17} have updated the stellar evolution models used in the BGM, to be able to simulate the seismic and chemical properties for all evolutionary stage. This new version of the BGM can also take into account the effect of extramixing occurring in giant stellar interiors. On the one hand, a first comparison with the large spectroscopic survey Gaia-ESO survey has been investigated in \cite[Lagarde \etal\ (2019)]{Lagarde19}, showing the importance of extra mixing to explain the carbon and nitrogen abundances observed at the surface of giant stars for different metallicity, masses, and ages (see Fig.\ref{fig2}). On the other hand, \cite[Cabral \etal\ (2019)]{Cabral19} coupled the BGM with a stoichiometric model to provide and study the chemical properties of planets building blocks in different stellar populations. \\

\paragraph{\textit{What about white dwarfs ? }}\\

To treat white dwarfs and their progenitors, we improve the BGM treatment of white dwarfs, including the cooling evolutionary models from Montr\'eal group and their corresponding synthetic colors (e.g.,  \cite[Fontaine \etal\ 2001]{Fontaine01}, and references therein , see Fig.\ref{fig1}). This improvement allows us to take into account pure C-core or C/O core white dwarfs models, as well as colors for DA and DB spectral types. The corresponding cooling models are computed for white dwarfs mass between 0.2 and 1.3 M$_{\odot}$, corresponding to an initial mass between $\sim$ 0.7 and 6.0 M$_{\odot}$ (using the initial-final mass relation from \cite[Cummings \etal\ 2018]{Cummings18}). In this new way to treat the white dwarf in the BGM, we also propose to generate white dwarfs in the Milky Way using the same initial-mass function and star formation rates as taken for all others stars. In a forthcoming paper, we plan to include cooling evolutionary models with more physics (e.g., mixing), different metallicities, and to be more consistent cooling models computed according to the evolutionary sequence of their progenitors. Those simulations will be compared with combined data from astrometric, asteroseismic and spectroscopic surveys. All these new improvements will be crucial to exploit the future large number of data concerning white dwarfs and to combine all of these different kinds of observations to constrain Galactic and stellar physics.


\begin{thebibliography}{}

\bibitem[Althaus et al. 2010]{Althaus10}
{Althaus LG, C\'orsico AH, Isern J, Garc\'ia-Berro E} 2010,
\textit{A\&A Rev.} 18, 471

\bibitem[Baglin \& Fridlund 2006]{CoRoT}
{Baglin, A., \& Fridlund, M.} 2006, 
\textit{in The CoRoT Mission Pre-Launch Status, Stellar Seismology and Planet Finding, eds. M. Fridlund, A. Baglin, J. Lochard, \& L. Conroy, ESA SP}, 1306, 11

\bibitem[Beck et al. 2012]{Beck12}
{Beck, P. G., Montalban, J., Kallinger, T., et al.} 2012, 
\textit{Nature}, 481, 55

\bibitem[Bedding et al. 2011]{Bedding11}
{Bedding, T. R., Mosser, B., Huber, D., et al.} 2011, 
\textit{Nature}, 471, 608

\bibitem[Cabral et al. (2019)]{Cabral19}
{Cabral, N, Lagarde, N., Reyl\'e, C., Guilbert-Lepoutre A. \& Robin, A.} 2019, 
\textit{A\&A}, 622, A49

\bibitem[Cummings et al. 2018]{Cummings18}
{Cummings J. D., Kalirai J. S., Tremblay P.-E., Ramirez-Ruiz E., Choi J.} 2018, 
\textit{ApJ}, 866, 21

\bibitem[C\'orsico et al. 2019]{Corsico19}
{C\'orsico, A. H., Althaus, L. G., Miller Bertolami, M. M., \& Kepler, S. O.} 2019,
\textit{A\&A Rev.}, 27, 7

\bibitem[Czekaj et al. 2014]{Czekaj14}
{Czekaj, M. A., Robin, A. C., Figueras, F., Luri, X., \& Haywood, M.} 2014, 
\textit{A\&A}, 564, A102

\bibitem[Eisenstein et al. 2006]{Eisenstein06}
{Eisenstein, D. J., Liebert, J., Koester, D., et al.} 2006b, 
\textit{AJ}, 132, 676 

\bibitem[Fontaine et al. (2001)]{Fontaine01}
{Fontaine, G, Brassard, P, \& Bergeron, P.} 2001, 
\textit{PASP}, 113, 409

\bibitem[Fontaine \& Brassard 2008]{FoBr08}
{Fontaine G, \& Brassard P} 2008,
\textit{PASP}, 120, 1043

\bibitem[Gaia Collaboration 2018]{GaiaDR2}
{Gaia Collaboration et al.}, 2018a, 
\textit{A\&A}, 616, A1

\bibitem[Gaia Collaboration 2018]{Babusiaux18}
{Gaia Collaboration, Babusiaux, C., van Leeuwen, F., et al.} 2018b, 
\textit{A\&A}, 616, A10

\bibitem[Genest-Beaulieu \& Bergeron 2014]{GeBe14}
{Genest-Beaulieu, C., \& Bergeron, P.} 2014, 
\textit{ApJ}, 796, 128

\bibitem[Gentile Fusillo et al. 2019]{Gentile2019}
{Gentile Fusillo, N. P., Tremblay, P.-E., G\"ansicke, B. T., et al.} 2019, 
\textit{MNRAS}, 482, 4570

\bibitem[Gilliland et al. 2010]{Kepler}
{Gilliland, R. L., Brown, T. M., Christensen-Dalsgaard, J., et al.} 2010, 
\textit{PASP}, 122, 131

\bibitem[Jordan 2007]{Jordan07}
{Jordan, S.} 2007, 
\textit{Astronomical Society of the Pacific
Conference Series}, Vol. 372, 15th European Workshop on
White Dwarfs, ed. R. Napiwotzki \& M. R. Burleigh, 139

\bibitem[Kepler et al. 2007]{Kepler07}
{Kepler, S. O., Kleinman, S. J., Nitta, A., et al.} 2007,
\textit{MNRAS}, 375, 1315

\bibitem[Kepler et al. 2016]{Kepler16}
{Kepler et al. 2016 Kepler, S. O., Pelisoli, I., Koester, D., et al. }2016, 
\textit{MNRAS}, 455, 3413

\bibitem[Koester \& Kepler]{KoKe15} 
{Koester, D., \& Kepler, S. O.} 2015, 
\textit{A\&A}, 583, A86

\bibitem[Lagarde \etal\ (2017)]{Lagarde17}
{Lagarde, N, Robin, A., Reyl\'e, C., Nasello, G.} 2017,
\textit{A\&A}, 601, A27

\bibitem[Lagarde \etal\ (2019)]{Lagarde19}
{Lagarde, N., Reyl\'e, C., Robin, A. C., et al.} 2019, 
\textit{A\&A}, 621, A24

\bibitem[Mosser et al. 2012a]{Mosser12a}
{Mosser, B., Goupil, M. J., Belkacem, K., et al.} 2012a, 
\textit{A\&A}, 548, A10

\bibitem[Mosser et al. 2012b]{Mosser12b}
{Mosser, B., Goupil, M. J., Belkacem, K., et al.} 2012b, 
\textit{A\&A}, 540, A143

\bibitem[Ricker et al., 2015]{TESS}
{Ricker G. R., et al.} 2015, 
\textit{Journal of Astronomical Telescopes, Instruments, and Systems}, 1, 014003

\bibitem[Robin et al. 2003]{Robin03}
{Robin, A. C., Reyl\'e, C., Derri\`ere, S., \& Picaud, S.} 2003, 
\textit{A\&A}, 409, 523

\bibitem[Stello et al. 2008]{Stello08}
{Stello, D., Bruntt, H., Preston, H., \& Buzasi, D.} 2008, 
\textit{ApJ}, 674, L53

\bibitem[Tremblay et al. 2011]{Tremblay11}
{Tremblay, P.-E., Bergeron, P., \& Gianninas, A.} 2011, 
\textit{ApJ}, 730, 128

\bibitem[Vauclair G (2013)]{Vauclair13}
{Vauclair G} 2013, 
\textit{Constraints on white dwarfs structure and evolution from
asteroseismology. In: Alecian G, Lebreton Y, Richard O, Vauclair G (eds)
EAS Publications Series, EAS Publications Series}, vol 63, pp 175?183

\bibitem[Vrard et al. 2016]{Vradr16}
{Vrard, M., Mosser, B., \& Samadi, R.} 2016, 
\textit{A\&A}, 588, A87

\bibitem[Winget DE \& Kepler SO 2008]{WiKe08}
{Winget DE \& Kepler SO} 2008,
\textit{ARA\&A}, 46, 157


\end{thebibliography}
\end{document}